\documentclass[10pt]{article}
\usepackage{graphicx} 

\usepackage[a4paper,top=2.5cm,bottom=2.5cm,left=2cm,right=2cm]{geometry} 
\usepackage[fontsize=13pt]{scrextend}
\usepackage{bbold}
\usepackage{authblk}
\usepackage{amssymb}
\usepackage{amsmath}
\usepackage{amsthm} 
\usepackage[dvipsnames]{xcolor}   

\usepackage{braket}
\usepackage{physics}
\usepackage{caption}
\usepackage{subcaption}
\usepackage{bbold}

\usepackage{graphicx}
\usepackage{listings}       
\usepackage{caption}
\usepackage{hyperref}     
\usepackage[normalem]{ulem}

\author[1]{Alessandro Civolani}
\author[1]{Vittoria Stanzione}
\author[1]{Maria Luisa Chiofalo}
\author[1,*]{Jorge Yago Malo}

\affil[1]{Dipartimento di Fisica Enrico Fermi, Università di Pisa and INFN, Largo B. Pontecorvo 3, I-56127 Pisa, Italy}

\affil[*]{jorge.yago@unipi.it}

\title{Engineering Transport via Collisional Noise: a Toolbox for Biology Systems}

\begin{document}

\maketitle
\abstract{The study of noise assisted transport in quantum systems is essential in a wide range of applications from near-term NISQ devices to models for quantum biology. Here, we study a generalised XXZ model in the presence of stochastic collision noise, which allows to describe environments beyond the standard Markovian formulation. Our analysis through the study of the local magnetization, the inverse participation ratio (IPR) or its generalisation, the Inverse Ergodicity Ratio (IER), showed clear regimes where the transport rate and coherence time can be controlled by the dissipation in a consistent manner. In addition, when considering several excitations, we characterize the interplay between collisions and system interactions identifying regimes in which transport is counterintuitively enhanced when increasing the collision rate, even in the case of initially separated excitations. These results constitute an example of the essential building blocks for the understanding of quantum transport in structured noisy and warm disordered environments.}

\section{Introduction}
Novel technological developments of quantum mechanical systems have allowed to include the effects of dissipative coupling to the environment providing not only a more realistic characterization of the hardware~\cite{Wiseman_Milburn,Petruccione_Breuer,Gardiner_Zoller}; but also, with the current level of control and tunability, giving access to novel non-equilibrium physical phenomena not appearing in closed systems: from the engineering of new states of matter~\cite{Daley_QJ_and_OQS,Harrington2022}, or the observation and description of dissipative phase transitions~\cite{Diehl2010,Kessler2012} to the control of dynamical rates~\cite{Vicentini2018,Li2018}, with examples in discrete~\cite{Skinner2019,Li2019,Bao2020} and continuous systems~\cite{Entanglement_free_fermion,Muller2022} among many other applications. In addition, the study of transport in noisy media has become essential for its potential applications in the description of biological systems. Following the discovery of long-coherence times at room temperature in complexes involved in photosynthesis \cite{Panitchayangkoon2010}, only possible with the development of femtosecond two-dimensional spectroscopy methods \cite{Fleming2011,Cao2020,Collini2010,Huelga2013}, a large community has initiated to question whether nature exploits the presence of coherent and dissipative couplings to enhance further efficiency \cite{OlayaCastro2008,Plenio2008}. 

In this context, we are interested in the phenomenological description of transport phenomena with the use of the toolbox provided by open quantum systems. In particular, we focus in bath descriptions that allow for flexible couplings and architectures and are compatible with noisy intermediate-scale quantum (NISQ) technology applications. As a result, we focus in the use of \textit{quantum collision models} or \textit{repeated interaction schemes} \cite{QCM_general_8, QCM_general_9,QCM_general_5,QCM_general_2, Collision_models_in_quantum_optics,QCM_general_picture_optics, QCM_general_thermometry,QCM_general_6,QCM_general_10,QCM_general_1,Quantum_collision_models_Open_system_dynamics_from_repeated_interactions}. This discrete bath description consists of itinerant degrees of freedom (d.o.f.) that interact instantaneously and one at a time with the system or a part of it. As a result, it is suitable for describing discrete lattice system or Quantum Computing (QC) devices. Importantly, it also bears strong similarities with biological systems were the collisions with ancillary systems can provide a good approximation to the noise present in those warm media. Thus, these models represent an efficient approach to building effective models for the description of such biological systems \cite{SCM_approach_to_transport,Modulation_of_Heat_Flux} that are also compatible with current NISQ devices for their simulation. 

Moreover, some of these applications have been more recently equipped with the toolbox of complex networks \cite{Biamonte2019}, creating a platform to study both the underlying microscopic transport phenomena together with the emergence of macroscopic properties, and leading to the development of a generalised theory composed of both quantum and classical tools \cite{DeDomenico2016}. The study of these networks and their transport properties, in their classical and quantum version, is referred to as quantum walks \cite{Kadian2021}. Dissipative versions of these problems, more recently developed, are being exploited to also tackle problems relevant to biology \cite{Rossi2017,Kurt2023}.

Inspired by these lines of research, in this work we consider a specific sub-branch of collision models, the \textit{stochastic collision models} (SCM)~\cite{Decoherence_without_entanglement_Quantum_Darwinism, SCM_approach_to_transport,Stochastic_versus_Periodic_Quantum_Collision_Models,Strategies_to_simulate_dephasing-assisted_quantum_transport_on_digital_quantum_computers,Stochastic_Collisional_Quantum_Thermometry}. In this formulation, the collisions are governed by a stochastic process and
the unraveling of the evolution of the system can be described in terms of individual random
realizations, which are tightly bound with individual runs on a given QC hardware.  As our aim is to consider the most general noise distribution, we take inspiration from \cite{SCM_approach_to_transport} and resort to the use of Weibull renewal processes \cite{Weibull_Renewal_Processes} as our stochastic distribution for a general and tunable noise description. In fact, this distribution allows to explore a variety of bath-induced noise regimes, by tailoring both the rate of collisions over time and their space and time homogeneity. This model allows to explore non-trivial structured baths via stochastic sampling \cite{SCM_approach_to_transport,Decoherence_without_entanglement_Quantum_Darwinism, Environment_qt} with a simple theoretical description comparable to other stochastic unravelings in standard Markovian dynamics~\cite{Daley_QJ_and_OQS}. 

Furthermore, quantum spin models have been used as buildings blocks for the study of these natural networks~\cite{Plenio2008} with architectures often based on the specific biological complex involved~\cite{SCM_approach_to_transport}. In our case, we consider instead an anisotropic linear chain with the idea of reducing the impact of the network topology and isolating the role of the dissipative coupling in the transport. In particular, in this work we investigate a paradigmatic quantum spin chain given by the Heisenberg XXZ model \cite{Heisenberg_original,Bethe_metals_original,Franchini_chains}, motivated by both its applicability to describe phenomenology in other biological systems, such as neuroscience of perception~\cite{YagoMalo2023}, and also, given its integrability, its historical use in the study of transport \cite{Prosen2011,Prosen_Buca_XXZ}. Thus, in this work we combine this quantum spin model with the SCM to investigate quantum transport of spin excitations and to access the interplay between coherent and dissipative dynamics also in the presence of interaction. 

The paper is structured in the following manner. In Section \ref{sec:model} we present the spin model and the details of the noise implementation, together with the state of the art in the understanding of transport slowdown via dissipation. In Section \ref{sec:results} we present the transport analysis while varying the noise parameters: we build up an understanding for the case of one single excitation, before extending to several interacting excitations. From our analysis, we show that noise can be engineered to control the system transport properties. While normal structure-less noise leads to transport slowdown \cite{Entanglement_free_fermion}, we find specific regimes where we can  minimize this effect, tailor the amount of coherent oscillations of the dynamics, or even enhance transport in the case where the particles are pinned together due to their interaction. Finally, in section \ref{sec:conclusion} we discuss our findings and potential future directions linking to disordered systems and complex networks.
\section{Model and Numerical Method}\label{sec:model}

\begin{figure}[tb]
  \centering
  \captionsetup{width=0.9\linewidth}
  \includegraphics[width=15.5cm]{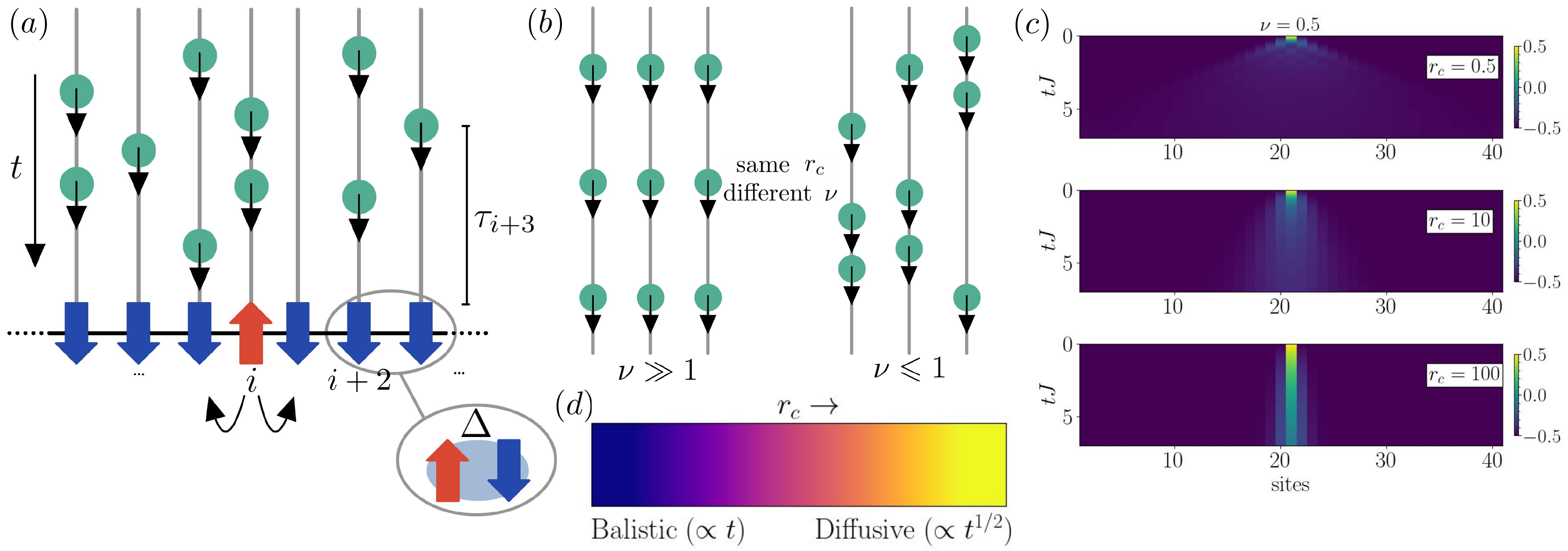}
    \caption{Concept of the model. (\textbf{a}) Diagram of an anisotropic spin chain with exchange rate $J$ and interaction strength $\Delta$ subject to stochastic collisional noise. Over time, individual ancillas (green circles) collide with individual spins in the chain with characteristic times $\tau_i$; (\textbf{b}) The noise distribution governing the collisions in Eq.~\ref{Weibull_noise} can be tuned to be strongly space and time heterogeneous. In particular, on the left-hand side we observe temporal and spatial homogeneity due to the high values of $\nu$, while on the right-hand side, for small values of $\nu$, we have strongly temporal heterogeneity; (\textbf{c}) Time evolution of the local magnetization $\langle \sigma^z_i (t)\rangle$, depicting the spreading of the initial excitation (spin defect at the central site) as we increase the collision rate $r_c=0.5,10,100$ (top to bottom) at fixed shape parameter $\nu=100$ (time homogeneous noise) for a system with $N=41$ spins. We observe that, similarly to the case of Markovian dephasing, the spreading velocity decreases with the increasing dissipative rate. (\textbf{d}) Phenomenology of transport slowdown due to dephasing between the ballistic regime ($r_c \ll 1$) where the system presents a linear spreading $(\propto\,t)$ and the diffusive regime ($r_c \ge 1)$ leading to slower propagation $(\propto\,t^{1/2})$, both appearing at long times/distances in Markovian dynamics.}
    \label{fig:summarizing_the_model}
\end{figure}

Here we consider an integrable generalisation of the Heisenberg spin chain that accounts for uni-axial anisotropy in the spin interaction, generally on the $z$-axis, described by the Hamiltonian \cite{Franchini_chains}:
\begin{equation}
    H_{XXZ}=J \sum_{i=1}^{N-1} \left[ \sigma^x_i \sigma^x_{i+1} + 
    \sigma^y_i \sigma^y_{i+1} + 
    \Delta \sigma^z_i \sigma^z_{i+1} \right]
    + h \sum_{i=1}^{N} \sigma^z_{i},
    \label{XXZ_Hamiltonian}
\end{equation}
with open boundary conditions (OBC). Here, $N$ represents the number of spins, $J$ is the exchange constant and the anisotropy is represented by the parameter $\Delta$ and $h$ representing a generic transversal field. This is schematically shown in Figure \ref{fig:summarizing_the_model}.
In the limit of moderate interaction $\Delta$, the sign of the coupling $J$, determines the ground state properties of the system which develops a \textit{ferromagnetic} ($J>0$) or \textit{anti-ferromagnetic} ($J<0$) order. Instead, when considering time-evolution or quench dynamics, we can see $J$ as the tunneling rate of an excitation or spin flip.

Unless stated otherwise, we work at fixed $J=1$ as our frequency unit and choose $h=0$, focusing on the case with $\Delta>0$ and making use of the following system's symmetry:

\begin{equation}
    \left[H_{XXZ}, S^z\right]=0,
    \label{XXZ_simmetry_conserved}
\end{equation}
meaning that the global magnetization along the $z$-axis, $S^z=\sum_i \sigma_i^z/2$, is preserved. 

In order to incorporate the effects of noise, we introduce the environment via a discretization, using the so-called quantum collision models \cite{History_QCM_1, History_QCM_2,History_QCM_3,History_QCM_4}, see Fig~\ref{fig:summarizing_the_model}a. In particular, we focus on one of their sub-branches: stochastic collision models \cite{SCM_approach_to_transport, Decoherence_without_entanglement_Quantum_Darwinism, QCM_quantum_Darwinism_1, QCM_quantum_Darwinism_3}. In these models, the collisions are governed by a stochastic process, and
the unraveling of the system’s evolution can be described in terms of single individual
realizations. In this type of framework, it is possible to introduce any probability distribution to describe the collisional rate. Since we are interested in understanding the role of the noise parameters in transport and the time evolution of the system, we consider, as in \cite{SCM_approach_to_transport}, a flexible distribution given by a Weibull
distribution \cite{Weibull_Renewal_Processes}:
\begin{equation}
    p(t)=\frac{\nu}{\mu}
    \left(
    \frac{t}{\mu}
    \right)^{\nu-1}
    e^{-(t/\mu)^\nu},
    \label{Weibull_noise}
\end{equation}

where $\nu \geq 0$, the shape parameter and $\mu > 0$, the scale parameter, are the collision parameters of the distribution. The shape parameter controls the temporal heterogeneity of the noise, (see Fig~\ref{fig:summarizing_the_model}b), describing heterogeneous collisions over time for $\nu \leq 1$ and temporal homogeneity for $\nu \gg 1$, as the intercollision time becomes constant.
 In contrast, the scale parameter is related to the overall collision rate that we define below:
\begin{equation}
    r_c=\frac{1}{\tau_{th}}=\frac{1}{\mu\:\Gamma\:(1+1/\nu)},
\end{equation}
with $\tau_{th}$ describing the mean collision time. As these rates can be tuned locally, changing the shape $\nu_i$ and scale $\mu_i$ parameters on each individual chain element can create not only temporal but also spatial heterogeneity in the system.

Now focusing on the evolution of the system, it is important to note that it is not possible to derive a GSKL-like master equation, as in the case of the Poisson distribution~\cite{Decoherence_without_entanglement_Quantum_Darwinism}, for a Weibull distribution given by Eq.~\eqref{Weibull_noise}. Despite this difficulty, it is still possible to describe the evolution in terms of quantum channels, see~\cite{SCM_approach_to_transport,Decoherence_without_entanglement_Quantum_Darwinism}, and relying on stochastic unraveling.

To do so, we initialized a list $\Bar{S}$ containing the first collision time of each site, randomly sampled from our distribution and extract from it $S_i=\mathrm{min}(S_j\:|\:\forall \:j\in[1,N]\:)$, the shortest waiting time before a collision. We then evolve the density matrix by computing its evolution until the time of the first collision, i.e. $\rho(S_i)=e^{-iH(S_i-t)}\, \rho(t) \,e^{iH(S_i-t)}$. At this time, we apply the quantum channel 
 \begin{equation}
     \Phi\left[\rho(S_i)\right]=\textrm{Tr}[ U_{\textrm{coll}}(\rho_a\otimes\rho(S_i))U^\dagger_{\textrm{coll}}]\,,
 \end{equation} 
 with $U_{\textrm{coll}}=\exp{-i(\pi/2)\sigma_a^x\otimes\sigma^z_i}$ representing the collision event and $S_i$ being the shortest waiting time before a collision. Note that here we have chosen the collisional event to produce dephasing in the system -- as it is proportional to $\sigma^z_i$ in the system. Thus, we are modelling a type of collision in which the interaction with the ancilla uncorrelates the system from other local d.o.f. while the same formalism can be adapted to other phenomena, e.g. collisions leading to the loss of the excitation ($\propto\,\sigma^-_i$).

After the computation of the first collision event, we can continue the process and redraw from the distribution in Eq.~\eqref{Weibull_noise} the next collision time for the $i$-th site, which will be $S_i=S_i+\tau$. We then update $\Bar{S}$ and extract the next collision time $S_i'$, and distinguish two situations: if $t+dt$, with $dt$ our numerical timestep, is larger than the new $S_i'$, we repeat the corresponding application of the new quantum channel; otherwise, we proceed with the time evolution by simply computing $\rho(S_i')=e^{-iH\:(S_i'-t)}\rho(S_i')e^{iH\:(S_i'-t)}$ and then apply the quantum channel. By iteratively repeating this process we can compute the time evolution of the system up to the desired time $T$.

\section{Results}\label{sec:results}

In this section, we present the results of our numerical analysis of the transport behaviour and relevant observables quantifying it. Since we consider OBC, we will refer to the final time of the simulations $t_f$ as the last time step at which the spreading of the magnetization is computed, given by the time at which the magnetization probability differs from its initial value in the boundary sites to minimize the finite size effects. After our numerical convergence analysis (see Appendix \ref{Convergence Analysis}), we concluded that choosing $dt = 0.02$ and the number of stochastic realizations $M = 500$ provide us with reasonable results in every noise situation. 
 
We start from an initial configuration with one(more) excitation(s), given by a flipped spin in a completely polarised state in the central site(s). Then we analyse the rate of spreading of the magnetization as a function of the system and noise parameters. We expect to observe that generally transport is slowed down with increasing the number of collisional events, as shown in Fig~\ref{fig:summarizing_the_model}c, so that we consider the limits of ballistic ($r_c \ll 1$) and diffusive ($r_c \ge 1)$ transport regimes, see Fig~\ref{fig:summarizing_the_model}d. While these have been observed in open systems and are well-characterized for structureless noises \cite{Entanglement_free_fermion} and for other related systems \cite{Eisler2011,Eichelkraut2013}, here, we want to generalise the study to more complex noise scenarios for the integrable XXZ chain.

\subsection{One excitation}

\begin{figure}[tb]
\centering
  \captionsetup{width=0.9\linewidth}
  \includegraphics[width=15.5cm]{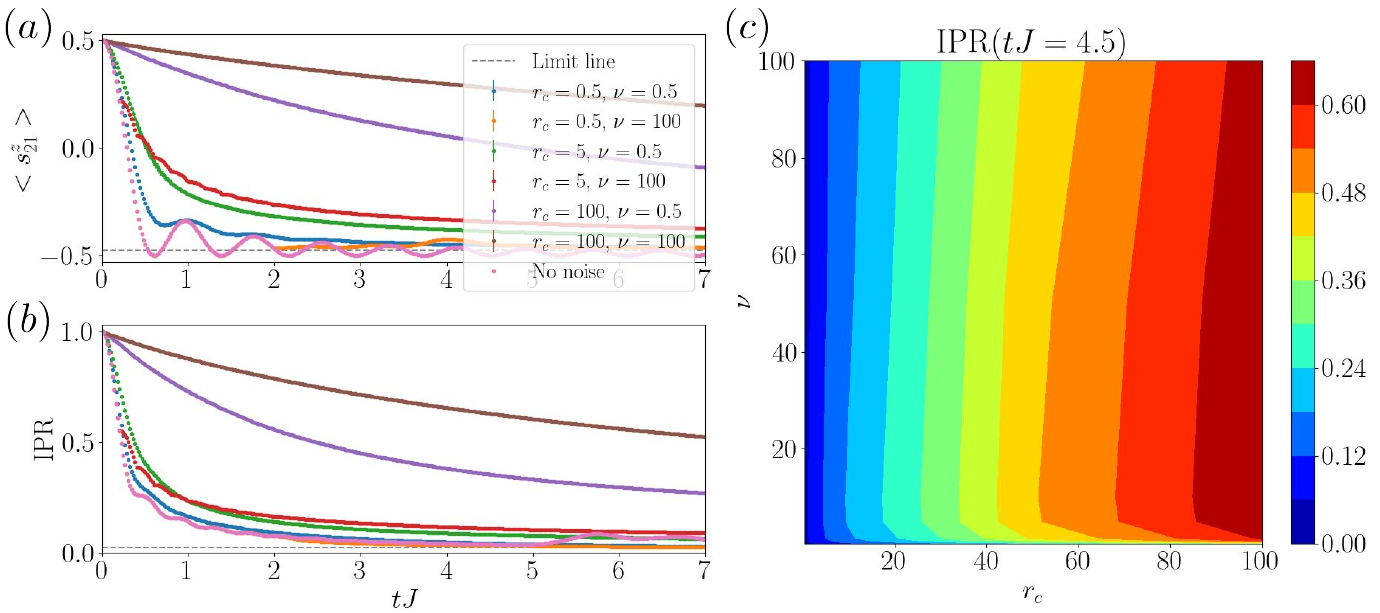}
    \caption{Transport of a single excitation. Magnetization and Inverse Participation Ratio (IPR) evolution in a system with 41 spin and $\Delta=0$. (\textbf{a}) Magnetization on the central site $\ev{s_{21}^z}=\ev{\sigma_{21}^z/2}$, where the excitation starts, versus time with varying collision parameters, $r_c$ and $\nu$.Notice that in general, increasing $r_c$ values correspond to decreasing transport speed. (\textbf{b}) IPR versus time with varied collision parameters, $r_c$ and $\nu$. We observe change in the transport rates with increasing $r_c$ and also reduced coherence times with smaller $\nu$. (\textbf{c}) IPR at time $tJ=4.5$ versus $\nu$ and $r_c$. Increasing $r_c$ leads to larger IPR (slower transport) apart from the limit at low $\nu$, we find that the larger effect occurs for shape parameters $\nu \sim 5$. Note that the errorbars are included in the plot but their size is comparable to the marker size. The simulation was performed with $dt=0.02$ and $M=500$ samples.}
    \label{fig:main_regimes}
\end{figure}

We now study how varying both our collision parameters $\nu$ and $r_c$ affects transport, as shown in Figure \ref{fig:main_regimes}. In general, as it would be the case in a standard Markovian regime, we see that increasing $r_c$ towards the diffusive regime decreases the transport speed. As the collisions are more frequent, the transport rate is reduced by the projective nature of the collisions, leading to a transport slowdown, which asymptotically leads to the quantum Zeno \cite{ Entanglement_free_fermion,Li2019} regime and the freezing of dynamics. By analysing the magnetization on the central site, see Fig~\ref{fig:main_regimes}{\textcolor{NavyBlue}{a}},  we observe that for time-heterogeneous noise (low $\nu$) the loss of coherence is significant and occurs already at $tJ\sim 2$. For noise uniformly distributed in time (high $\nu$), instead, the oscillatory behaviour highlights that some level of coherence is preserved, especially in the low-noise limit (low $r_c$) for the studied times. Thus, we can conclude that the homogeneous regime, closer to the Markovian limit, preserves stronger levels of coherence in the system (for equal $r_c$) due to the fact that collisions occur on average in orderly time intervals.

In addition to analyse the local magnetization profiles, we quantify the delocalization behaviour and transport rate making use of the inverse participation ratio (IPR) \cite{CTQW_spatially_correlated_noisy}. The IPR is: 
\begin{equation}
    \mathrm{IPR}= \sum_{i=1}^N \bra{i}\rho(t)\ket{i}^2, 
    \label{IPR_formula}
\end{equation}
where $i$ represents the site index and $\ket{i}$  represents the single-excitation localized states  $  \Big\{
    \ket{i}\: : \:\ket{i}=\sigma_{i}^{\dagger} \ket{0}^{\otimes N}
    \Big\} $ \cite{SCM_approach_to_transport}. Thus, IPR is bounded between the complete delocalization asymptotic value $\mathrm{IPR}=1/N$, and $\mathrm{IPR}=1$, corresponding to complete localization, i.e. when the excitation remains on a particular site of the network \cite{CTQW_spatially_correlated_noisy}. Therefore, the larger the $\mathrm{IPR}$, the more localized the excitation is over the lattice. We generally expect that, in the absence of dissipation, long-lasting Bloch oscillations set in, while in the long-time limit in the presence of dephasing, the excitation spreads over the lattice and coherence dissapears tending to a configuration where the excitation is completely delocalized over the chain. Thus, the rate at which we approach this state -- and consequently the evolution of the IPR -- strongly depends on the noise parameters.

\begin{figure}[tb]
\centering
  \captionsetup{width=0.9\linewidth}
  \includegraphics[width=15.5cm]{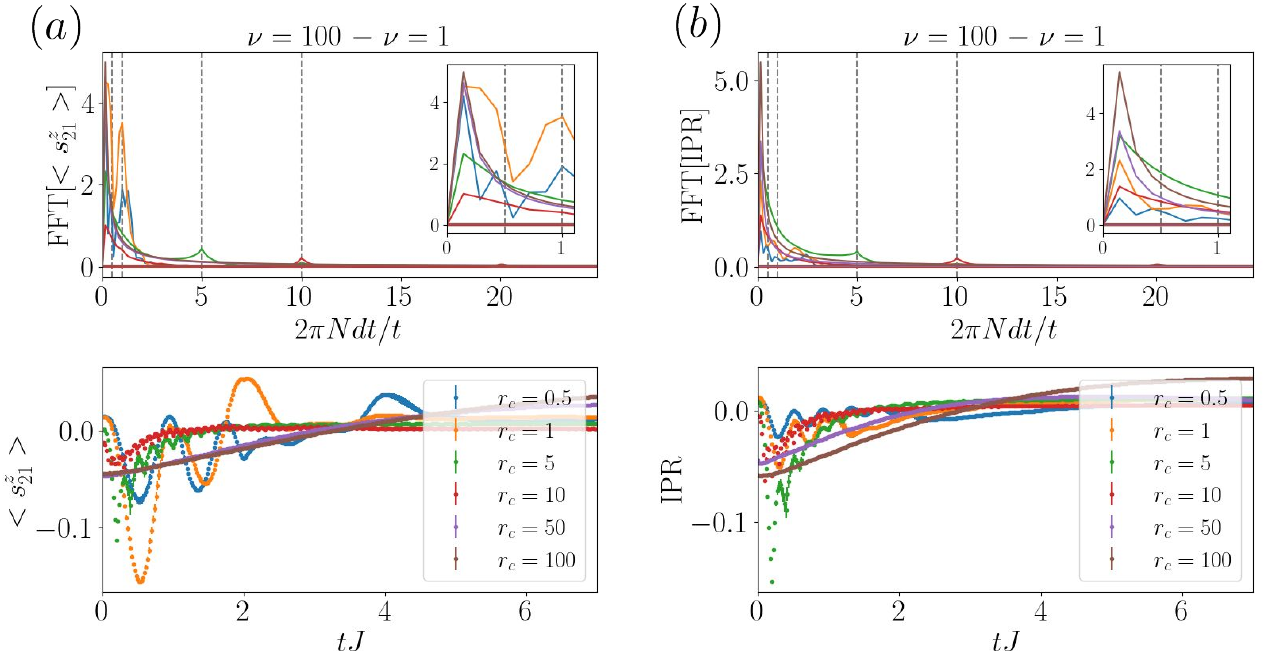}
    \caption{\textit{Fast Fourier Transform} analysis for the single excitation case. Both of the time signals analysed have been obtained by subtracting the curves corresponding to $\nu=100$ and $\nu=1$ at equal $r_c$ and we have removed the zero frequency component. The results correspond to a system with 41 sites, 1 excitation and $\Delta=0$. (\textbf{a}) FFT of the central magnetization signal (top panel) and the signal itself (bottom panel). We observe that the main peak corresponds with the respective $r_c$ values apart from some spurious effects at low frequency. (\textbf{b}) FFT of the IPR signal (top panel) and the signal itself (bottom panel). We confirm the presence of the largest peak at $r_c$. The insets correspond to the both spectra at low frequencies to facilitate visualization.
 }
    \label{fig:FFT_analysis}
\end{figure}

In Fig~\ref{fig:main_regimes}{b}, we provide the analysis of the IPR behaviour with varied noise parameters and we identified different regimes. For fixed temporal heterogeneity with $\nu = 0.5$, we observe that, in general, increasing $r_c$ implies a slowdown in the transport as it is the case of structureless noise and a quick damping of the oscillatory behaviour. Note that the IPR is less sensible to coherence than the local magnetization due to the fact that it is a lattice-averaged measure. 
From the analysis of both magnetization and IPR we can see that in the large $r_c = 100$ and low $\nu$ regime, we have groups or bunches of ancillas interacting with the system in short intervals of time. In these cases, the effect of subsequent ancillas is small, given that the first one has already projected and locally disentangled the system and time is required for the system to rebuild entanglement. Thus, this regime presents a lower effective $r_c$ and smaller noise sensitivity. In fact that can be confirmed if comparing datasets with the same $r_c$ but different $\nu$. The case of uniformly distributed noise in time ($\nu =100$), with collisions occurring on average at the same rate in each location even within individual realizations, has a notably smaller impact on the system coherence. Note how these oscillatory behaviours have a frequency dependence with the value of $r_c$, as we will discuss in detail in Figure \ref{fig:FFT_analysis}.

If we analyse the results from fixed collision rates $r_c$, for low collision rates ($r_c\sim0.5, 1$) we observe a moderate transport slowdown, less evident for smaller shape parameters, as discussed  before. For high-noise presence instead, we study up to $r_c\sim100$, transport suffers a heavy slowdown. This effect is a manifestation of the quantum Zeno effect \cite{Li2019}, leading to the freezing of the transport.

We provide a summary of this behaviour report in Fig~\ref{fig:main_regimes}{c}, where we display the IPR at a given intermediate time ($tJ=4.5$) versus $\nu$ and $r_c$. There we observe how: i) the increase of the collisional rate $r_c$ leads to general transport slowdown; ii) while the shape parameters has a small impact in the transport rate, we observe clearly the change in trend due to the effects at $\nu\ll 1$ where the effective $r_c$ is smaller (leading to smaller values of IPR). We also find that for $\nu~5$, with the specific value being a function of $r_c$, corresponds to the noise with larger impact on transport, with a slow decrease towards $\nu \gg 1$. This shows how by tuning both $\nu$ and $r_c$ there is a high degree of control on the transport with more tunability if we consider local site independent changes. We leave this possibility to future studies.

In order to understand the coherent behaviour observed in the time signals, in Figure \ref{fig:FFT_analysis}, we analyse the spectrum via the \textit{Fast Fourier Transform} (FFT) of both the local magnetization time evolution (Fig \ref{fig:FFT_analysis}a) and the IPR time evolution (Fig \ref{fig:FFT_analysis}b), to characterize the oscillatory effects. In particular, in our analysis we subtract the time heterogeneous $(\nu=1)$ from the time homogeneous $(\nu=100)$ signal for the same collisional rate $r_c$ and search for the main frequencies of the system, removing the DC component. From this procedure we observe coherence-driven oscillatory behaviours in the system, with frequencies that vary with $r_c$ and seemingly independent of $\nu$. As we increase $r_c$, oscillations are washed out by the lack of coherence and the strong damping of the dynamics the main peak, always at frequency $\sim r_c$, becomes smaller. However, this remains as the main frequency apart from the DC component and some spurious effects at low frequencies.

\subsection{Multiple excitations}

\begin{figure}[tb!]
  \captionsetup{width=0.9\linewidth}
  \includegraphics[width=15.5cm]{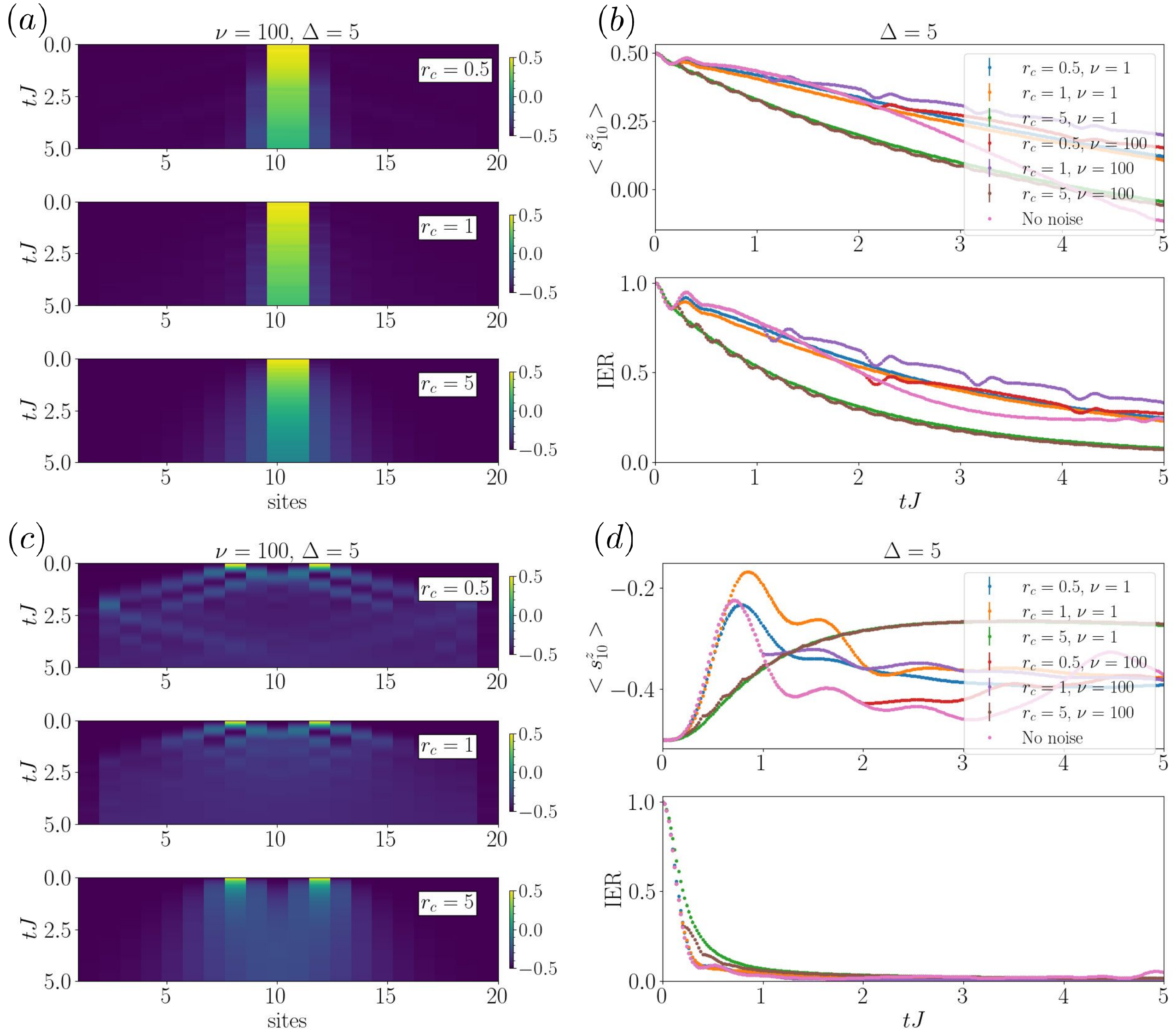}
    \caption{ Multiple excitation case. (\textbf{a}) Spreading of neighbouring excitations with fixed $\nu=100$. Here, we consider an initial state in which the excitations are located in the middle sites, with the magnetization spreading over the chain with different rates depending on $r_c$.
     (\textbf{b}) Evolution of the magnetization (top panel) and of the Inverse Ergodicity Ratio (IER) (bottom panel) versus time for different noise collisional rates $r_c$ and shape parameters $\nu$ as in the legend for two neighbouring excitations. We observe that noise prevents the pinning of the excitations. For the case of homogeneous noise, frequency-dependent modulations appear.(\textbf{c})-(\textbf{d}) Same as (\textbf{a})-(\textbf{b}) with the two excitations initially separated by three spins. From the transport side, we observe small changes with $r_c$. However, we find that noise can prevent coherent destructive interference enhancing transport in localized regions of the chain. This simulation was performed in the 20-sites lattice, with $\Delta=5$, and varying the collision rate $r_c=0.5,1,5$. The numerical parameters correspond to $dt=0.02$ and $M=250$ samples.
  }\label{fig:2exc_strong_anisotropy}
\end{figure}

So far we have analysed the role of noise in the transport but not yet considered its interplay with other competing physical effects. Thus, we now add the effect of the anisotropy $\Delta$ to our study and consider how the interaction between excitations can affect the transport in the system. 

Given the exponential growth of the Hilbert space $\mathcal{H}$, even within restricted magnetization sectors, and our need to average over trajectories, which involve the evolution of the operator $\rho$- whose dimension is $(\text{dim} \mathcal{H})^2$ -  we reduced the size of the system when studying more than one excitation. The smaller chain also exhibited faster convergence, allowing to reduce the required trajectories. Furthermore, given that we do not expect new phenomena to appear for large $r_c$ where the Zeno regime dominates, we also restricted the study to moderate collisional rates $r_c \leq 5$. Thus, we study the magnetization, its spreading and related observables for different noise structures after varying the anisotropy.

It is important to note that the IPR is defined in the case of a single excitation. Thus, here we need to generalise IPR to the case of a generic number of excitations. Several extensions to the IPR have been discussed in literature, through e.g. the introduction of the Generalized Inverse Participation Ratio \cite{Generalized_IPR}.  However, this quantity is computationally costly and does not take advantage of the structure of our simulation. 

Instead, we chose to introduce a new quantity, that we denote \textit{Inverse Ergodicity Ratio} (IER):
 \begin{equation}
     \text{IER(t)}= \sum_j \bra{j} \rho(t) \ket{j},
 \end{equation}
defined in terms of the multiple-excitation states $\ket{j}$ that compose our computational basis. In fact, this can be seen as a Fock basis for the occupation number:
  $  \Big\{
    \ket{j}\: : \:\ket{j}=\sigma_{i_1}^+\sigma_{i_2}^+
    \dots\sigma_{i_{q-1}}^+\sigma_{i_q}^+\ket{0}^{\otimes N}
    \Big\} $, where the subscripts $i_l$ indicate the sites in which we have the  $q$ excitations and the state index $j$ runs from 1 to $d=\text{dim}\mathcal{H}$.  

The IER bounds, now, describe the following limiting behaviours: $\mathrm{IER}=1/\text{dim} \mathcal{H}$ implies that the system is in a superposition of all the possible states in our reduced magnetization sector basis, thus the state is \textit{ergodic} \cite{ET_hypothesis}; $\mathrm{IER}=1$ instead  refers to the case in which the system is one out of these specific configurations, that forms part of our basis. So, in analogy with the IPR, the states that are more delocalized (low IPR) are also the more ergodic ones (low IER).

Focusing on the study of the model with anisotropy $\Delta$, we have analysed all the regimes of noise structures and $\Delta$. We here report on those that manifested interesting properties, illustrated in Figure \ref{fig:2exc_strong_anisotropy}, for the simpler case of two neighbouring excitations. In particular, we show the spreading of magnetization, the evolution of the magnetization and of the IER versus time varying the two collision parameters $r_c$ and $\nu$ for two excitations located in the middle of the chain (Fig~ \ref{fig:2exc_strong_anisotropy}a and Fig~ \ref{fig:2exc_strong_anisotropy}b). In contrast with the case of one single excitation, where no regimes of enhanced transport could be found, here in the presence of large anisotropy $\Delta$, increasing collision rates lead to sustained transport, with the noise breaking the pinning of the excitations together. This important result is supported also by the analysis of the case of weaker $\Delta=2.5$, shown in Appendix \ref{append.case_delta2.5}, where the effect is though less evident, still present. We also observe some periodic effects appear at high $\nu$ (see the $\nu=100$ case in Fig~\ref{fig:2exc_strong_anisotropy}b) with the main frequency component at $r_c$.

Since spatial separation of the excitations would drastically reduce their interactions, we consider the generality and robustness of these results by placing them at a fixed distance initially (see Fig~ \ref{fig:2exc_strong_anisotropy}c and \ref{fig:2exc_strong_anisotropy}d). In Fig~\ref{fig:2exc_strong_anisotropy}c, we observe that the noise can prevent certain destructive interference that appears due to coherence in the absence of noise (or small $r_c$). In this case, noise enhances spreading in the central part of the system. This could be useful in engineered irregular architectures beyond linear chains to prevent interference effects. By analyzing the magnetization and IER in Fig~ \ref{fig:2exc_strong_anisotropy}d, no important differences are observed by increasing $r_c$, apart from a small slowdown of the spreading, in fact regaining a similar behaviour of the single excitation case but with smaller impact from the noise. 
We note that, unlike the case with two neighbouring excitations, where the transport was mostly prevented due to the anisotropy, here we observe differences with $r_c$. In particular, an increase in the collision rate $r_c$ quite rapidly breaks the coherence-driven oscillations, and mostly independently of the noise time-homogeneity.


\section{Conclusion}\label{sec:conclusion}

In this work, we have presented a simple framework for the study of transport modulation via dissipation with relevant applications not only in quantum technologies but also in phenomenological studies for biology.  We have explored how systematic tuning over the noise parameters acting on our physical system can modify, in a controllable manner, its transport properties. This has been possible by exploiting a recently developed method to describe open quantum systems, that is based on \textit{stochastic collision models} (SCMs). This description allowed us to tailor the noise, by tuning both the number of collisions occurring over time and their time homogeneity. In so doing, we could access a large variety of bath-induced noise regimes within a sustainable numerical approach. In particular, our analysis of the noisy XXZ chain concluded that even if structure-less noise leads to the unavoidable slowdown of dynamics \cite{Entanglement_free_fermion}, here it is possible to find regimes where the transport and coherence time can be controlled by the dissipation in a consistent way. 

More importantly, when considering several excitations, we found an interesting interplay between collisions and system interactions. From this outcome, we identified regimes in which transport is enhanced when increasing the collision rate. We could also highlight differences in the transport rate within certain parts of the system, when the particles were not initially next to each other. This is a result of the controlled destruction of coherence that induces destructive interference which would hamper the transport in part of the system. All these results contrast the single excitation or low-anisotropy cases in which, in general, the higher is presence of noise, the slower becomes the transport rate. In addition, since we observed that noise allows for a quicker spreading in the central part of the system, this mechanism could prove useful when considering more complex lattice topologies.

All in all, our results suggest that it is possible to find so-far unexplored scenarios where the noise can assist the system transport. This can be especially useful in the presence of other competing mechanisms that prevent transport or with more complex connectivities. Thus, our results can be relevant in the investigation of disordered systems \cite{MBL_Nandkishore}, so far studied also from the one-particle perspective through IPR \cite{MBL_from_IPR_study} and in the context of open quantum systems \cite{Jorge_Localization}, as well as in randomly connected networks \cite{SCM_approach_to_transport} relevant to biology problems, where we could enhance and control transport by the engineering of collision models. We believe our study constitutes one of first step in building this understanding, and fostering this exploration.

\subsection*{Acknowledgement}
The authors would like to thank A. Daley, G. M. Cicchini and M. C. Morrone for inspiring and useful discussions. V. Stanzione would like to thank C.V. Stanzione for useful suggestions for collecting the data. V.S was supported by the project PNRR - HPC, Big Data and Quantum Computing – CN1 Spoke 10, CUP I53C22000690001. J.Y.M. was supported by the European Social Fund REACT EU through the Italian national program PON 2014-2020, DM MUR 1062/2021. M.L.C. acknowledges support from the MIT-UNIPI program and by the National
Quantum Science and Technology Institute (NQSTI),
spokes 2 and 10, funded under the National Recovery and Resilience Plan (NRRP), Mission 4 Component 2 Investment 1.3 - Call for tender No. 341 of 15/03/2022 of Italian Ministry of University and Research, funded by
the European Union NextGenerationEU, award number
PE0000023, Concession Decree No. 1564 of 11/10/2022
adopted by the Italian Ministry of University and Research, CUP D93C22000940001 and from the Spoke 10-HPC, Big Data and Quantum Computing - CN00000013 - CUP I53C22000690001.

\bibliography{Biblio}
\bibliographystyle{unsrt}


\vspace{6pt} 

\appendix
\section{Convergence Analysis}
\label{Convergence Analysis}

In this section we summarise the numerical details of our simulations. We also provide some examples of the relevant numerical analysis that preceded our study to ensure that our results were computationally sound and method-independent.

First, it is important to note the choice for the final time $t_f$ for which the spreading of the magnetization is computed in our simulations -- for the given noise structure analysed or for the one with faster spreading, if we are considering a comparison. We choose this time $t_f$ taking into account that our chain is finite and we require that the excitation has not already reached the boundary sites to prevent the introduction of boundary effects. Then, unless otherwise specified, the evolution is performed until the time step at which the magnetization over the two boundary sites remains completely polarised down as set at $t=0$.

Regarding the choice of $dt$, the presence of noise, causes our observables to depend on the particular trajectory, so our numerical parameter analysis is taken in terms of the comparison of averages over samples with the desired parameters to be tested. To verify the regime of convergence for both the timestep $dt$ and the number of samples $M$ with varied noise structures we chose the same system size of $N=41$ sites as in the main text. While this system size is not limiting given the fixed magnetization of our model, in longer chains the effects of the collisions would be more difficult to be observed given the \textit{diluteness} with a single excitation for most observables.
Here, we analyse the magnetization on the central site because it is more subject to numerical parameters changes than lattice-averaged quantities such as IPR.

\begin{figure}[tb]
  \centering
  \captionsetup{width=0.9\linewidth}
  \includegraphics[width=14cm]{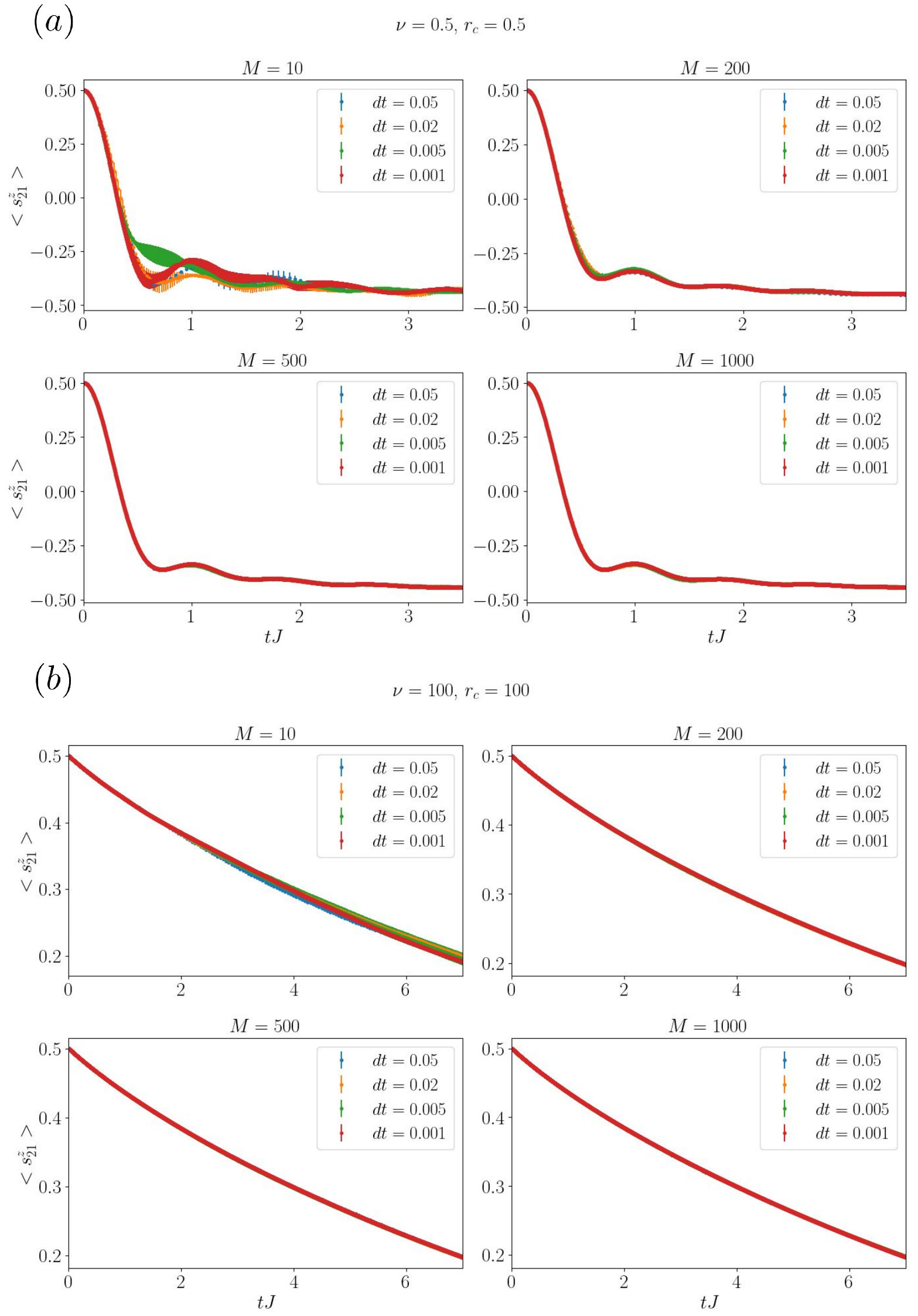}
  \caption{Benchmarking the code. (\textbf{a}) Evolution of the magnetization at the central site for a 41-sites chain and one single excitation in the middle, with $\nu=0.5$, $r_c=0.5$, $\Delta =0$.
Each plot refers to a fixed  $M=10$, 200, 500, 1000, and different time steps $dt=0.02$, 0.01, 0.005, 0.001, as in the legends. (\textbf{b}) Evolution of the magnetization $s_{21}^z$ on central site for the case of 41 sites, a single excitation in the middle and $\nu=100$, $r_c=100$, $\Delta =0$. Plot obtained changing the sample size $dt=0.02$, 0.01, 0.005, 0.001 at fixed $M=10$, 200, 500, 1000.}
\label{fig:Benchmarking_the_code}
\end{figure}

In Fig~\ref{fig:Benchmarking_the_code}a, we compare the magnetization on the central site $\ev{s^z_{21}}$, at given $M$ values and for different time steps $dt$, in one of the limiting cases of the noise parameters with $\nu=0.5$, $r_c=0.5$ and $\Delta =0$. In Fig~\ref{fig:Benchmarking_the_code}b, we then report the opposite limit the cases of different noise structures: $\nu=100$, $r_c=100$, $\Delta =0$. Please note that only these two example convergence plots are reported for the sake of conciseness but the procedure was repeated for each noise regime.

As we can observe, the averaged observables obtained converge already within $M=500$ (we provide 1000 for comparison) as they overlap within errorbars for every $dt$ and for both the noise structures analysed. Thus, guaranteeing the choise of $M=500$.
 Finally, we observe good convergence for all the $dt$ included and we choose an intermediate value $dt=0.02$ for numerical optimization.

\section{Case $\Delta = 2.5$}\label{append.case_delta2.5}

\begin{figure}[tb]
  \centering
  \captionsetup{width=0.9\linewidth}
  \includegraphics[width=15.5cm]{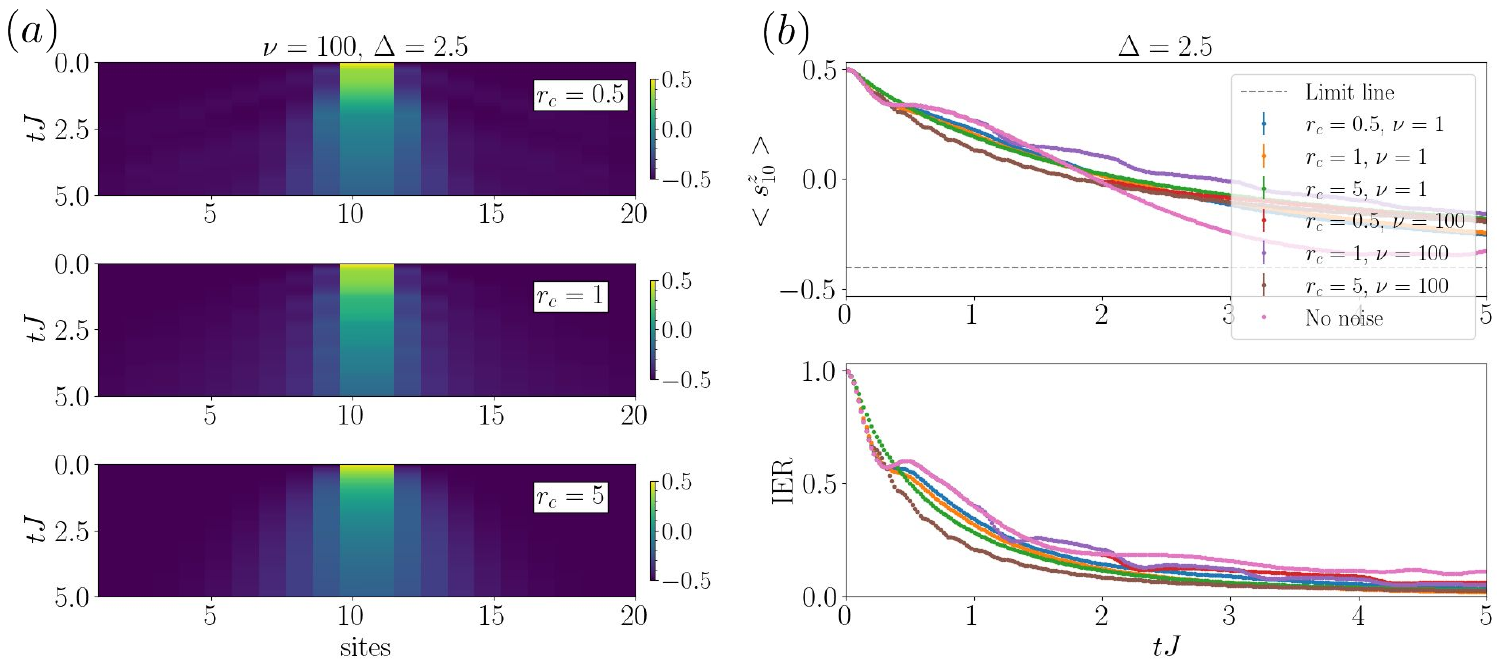}
  \caption{Case of two excitations and $\Delta=2.5$. (\textbf{a}) Spreading of the excitation varying the collision rate $r_c=0.5,1,5$ at fixed interaction strength $\Delta=2.5$ and different shape parameter $\nu=1$, 100. Here, we consider an initial state in which the excitations are initially in the middle sites, the magnetization spreads over the chain with different speeds depending on $r_c$. Note how the coherent effects are visible only for the case of $\nu=100$ and however, the trends apart from the oscillations, are very similar between $\nu=1$ and $\nu=100$. These effects are more visible in the case of $\Delta=5$ reported in Figure \ref{fig:2exc_strong_anisotropy}.
  (\textbf{b}) Evolution of the magnetization and IER for the case of 20 sites, 2 excitations, $\Delta=2.5$, and fixed $\nu=100$. In general, increasing $r_c$ produces a slowdown in transport. Unlike the case of a single excitation, here with large values of anisotropy $\Delta$ increasing the collision rate supports the transport as the noise helps breaking the pinning of the excitations together. In addition, we can observe that, again, for noise evenly distributed over time, some frequency-dependent modulations appear. The simulation was performed with $dt=0.02$ and 250 samples.}
    \label{fig:Analysis_2exc_Delta=2.5}
\end{figure}

In Figure \ref{fig:Analysis_2exc_Delta=2.5} we include the results for the case of $\Delta=2.5$ varying $\nu$ and $r_c$ for the same parameters included in Figure \ref{fig:2exc_strong_anisotropy} discussed in the Result section. This a complementary analysis to the results for the case of $\Delta=5$. We observe that the results for moderate $\Delta$ support our findings, in which we can identify regimes for enhanced transport when we increase the amount of collisions even if the effects are now smaller.


\end{document}